\documentclass[%
10pt,
superscriptaddress,
aps,
prl,
nofootinbib,
nobibnotes,
amsmath,
amssymb,
twocolumn,
floatfix,
]{revtex4-2}
\usepackage{amssymb}
\usepackage{placeins}
\usepackage{physics}
\usepackage{graphicx}
\usepackage{import}
\usepackage{amsmath}
\usepackage{comment}
\usepackage{amsthm}
\usepackage{amsfonts}
\usepackage[normalem]{ulem}

\usepackage{xcolor}
\usepackage[colorlinks]{hyperref}
\usepackage[caption=false, subrefformat=parens]{subfig}

\newcommand{\dphi}{S}
\newcommand{\Kk}{K {k^*}^2}
\newcommand\Keff{K_\mathrm{eff}}
\newcommand\ZZ{\mathbb{Z}}
\newcommand\rO{\mathrm{O}}

\newcommand\rU{\mathrm{U}}

\let\oldsection\section
\renewcommand\section[1]{\emph{#1}.---}

\begin{document}

\title{Ising criticality can drive vortex deconfinement in a spin-orbit coupled Bose gas}

\author{Stuart Yi-Thomas}
\email[]{snthomas@umd.edu}
\affiliation{Condensed Matter Theory Center and Joint Quantum Institute, Department of Physics, University of Maryland, College Park, Maryland 20742-4111, USA}
\author{David M. Long}
\affiliation{Condensed Matter Theory Center and Joint Quantum Institute, Department of Physics, University of Maryland, College Park, Maryland 20742-4111, USA}
\affiliation{Department of Physics, Stanford University, Stanford, California 94305, USA}
\author{Jay D. Sau}
\affiliation{Condensed Matter Theory Center and Joint Quantum Institute, Department of Physics, University of Maryland, College Park, Maryland 20742-4111, USA}

\date{March 16, 2026}

\begin{abstract}
Spin-orbit coupling in Bose gases is known to lead to an Ising-symmetry--broken phase where the bosons condense at one of two nonzero momenta. In two dimensions, the finite momentum of the order parameter allows vortex-antivortex pairs that are typically bound in the superfluid phase to freely separate along Ising domain walls. This non-trivial interaction between the superfluid and the Ising order suggests that the critical fluctuations near an Ising transition could drive a Berezinskii-Kosterlitz-Thouless transition of the superfluid. We present numerical evidence of this phenomenon using a Monte Carlo simulation that shows the disappearance of superfluid stiffness near an Ising transition. Additionally, we find numerical evidence that the Ising phase transition becomes first order and we justify this claim with a variational approximation.
\end{abstract}

\maketitle

The exquisite experimental control available in ultracold bosonic gases allows the emulation of a wide variety of physical models.
Some of these are natural models of solid state phenomena, but others have no known analogue in material physics.
For instance, several methods realize bosonic models with a ``double-well dispersion'' relation~\cite{parker2013,clark2016,zheng2014,lin2011,zhai2015,galitski2013}, analogous to spin-orbit coupling of electrons in solid state systems.
However, unlike electrons, bosons with such a dispersion condense into one of the two wells at low temperatures.
In addition to the superfluid condensation transition, there can be an Ising transition separating a reflection-symmetric phase from a symmetry broken phase, where all bosons condense into one well and not the other~\cite{lin2011}.

Though experiments on these double-well-dispersing bosons are not natively one-dimensional~\cite{clark2016}, theoretical studies have largely been restricted to this case. Even here, the interplay between the superfluid order and the Ising order constrains the motion of Ising domain walls~\cite{liu2016exotic,bhattacharjee2025}, leading to Ohmic transport~\cite{chou2024}.
However, neither the Ising symmetry nor the $\rU(1)$ superfluid have long range order in one dimension, and are therefore defined only in the sense of crossovers.
Interestingly, quantum phase transitions of the Lifshitz variety are theoretically predicted to appear in these systems~\cite{sachdev1996zero,yang2004ferromagnetic,cole2019},
which can be expected based on the association of quantum phase transitions with classical transitions in one higher dimension. Later analysis has shown, however, that quantum fluctuations drive this transition to be first order~\cite{kozii2017}.

\begin{figure}[t]
\centering
\includegraphics[width=1.0\columnwidth]{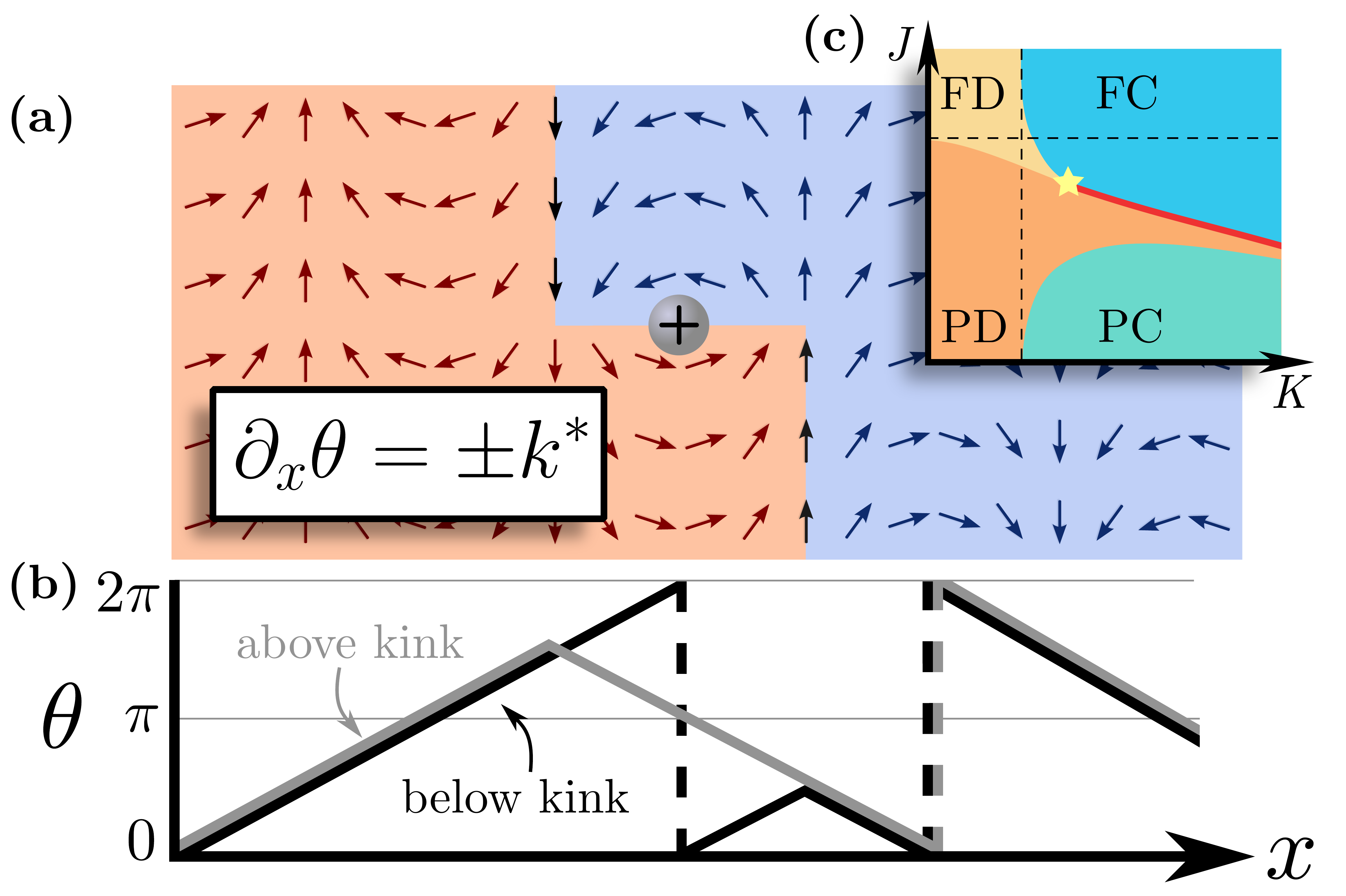}
\caption{\label{fig:cartoon} \textbf{(a)} Cartoon schematic of a deconfined vortex on a kink in a vertical domain wall between vortex-confined states with different momenta $k^*$. 
As this kink is a local defect, its energy is independent of system size, and it does not experience a long-range force from other kinks. 
\textbf{(b)} Plot of the twisting of the phase $\theta$ above and below the kink. 
The dotted line shows the wrapping from $2\pi$ to $0$.
\textbf{(c)} Schematic phase diagram of the model in Eq.~\ref{eq:lattice-action} as a function of Ising coupling $J$ and XY coupling $K$ demonstrating four phases, denoted by ferromagnetic/paramagnetic (F/P) and vortex confined/deconfined (C/D). The Ising transition leads to a stiffness collapse in the XY model. The red line denotes a first order transition and the star denotes a purported tricritical point.
The dashed lines represent decoupled ($k^*=0$) phase boundaries.
}

\end{figure}

In contrast, two dimensional double-well dispersing Bose gases demonstrate both a true Ising transition and a Berezinskii–Kosterlitz–Thouless (BKT) vortex confinement transition.
In this Letter, we study the relationship between the Ising phase transition and the BKT transition. 
We start by showing that the Ising and \(\rU(1)\) superfluid symmetries associated with these transitions are not independent, but rather form a semidirect product.
We find that this semidirect structure leads to a deconfinement of vortices along domain walls of the Ising order parameter. 
In the vicinity of an Ising transition, where large domain walls percolate through the system, the deconfinement of vortices leads to the collapse of the superfluid stiffness. Thus, there cannot be a direct transition between Ising phases while vortices remain confined, as shown in Fig.~\ref{fig:cartoon}(c).
We confirm this prediction through Monte Carlo numerics and a variational calculation, and additionally find that the Ising transition becomes first order.

\section{Interplay of symmetry defects}%
The symmetry structure of the double-well dispersing system suggests an interaction between Ising domain walls and BKT vortices.
The \(\ZZ_2\) Ising symmetry is a reflection of the \(\rU(1)\) phase of the boson, which has a nontrivial effect on the $\rU(1)$ phase rotation symmetry.
The global symmetry group is the semidirect product $\rU(1) \rtimes \ZZ_2$, isomorphic to $\rO(2)$.
Our key observation is that 
certain Ising domain walls bind \(\rU(1)\) vortices as a consequence of this non-Abelian total symmetry group.
The \(\ZZ_2\) symmetry-broken states are characterized by a gradient in the phase of the superfluid order parameter corresponding to the momentum of one of the two minima in the dispersion relation, \(k= \pm k^*\).
Integrating this phase twist around a domain wall parallel to $k^*$ (shown as horizontal in Fig.~\ref{fig:cartoon}) reveals that they carry
a linear density $|k^*|/\pi$ of bound vortices. 
Therefore, a horizontal kink of width $\pi/|k^*|$ in a vertical domain wall binds a single vortex (see Fig.~\ref{fig:cartoon}).
The energy cost of the vortex kink on top of the vertical domain wall is intensive, and the vortex is free to move along the domain wall.
Thus, near the Ising transition, where domain walls percolate through space, vortices can bind to these domain walls and deconfine.

This analysis of interrelated symmetry defects suggests that the Ising transition should always be accompanied by a vortex deconfinement transition. Consequently, there can be no direct transition between the phase in which all the bosons condense in a single well and one in which half the bosons condense in each well [Fig.~\ref{fig:cartoon}(c)].

\section{Monte Carlo calculation}%
To numerically verify the predicted phase diagram in Fig.~\ref{fig:cartoon}(c), we define a minimal lattice model with the same $\rU(1) \rtimes \ZZ_2$ symmetry as the Bose gas, but which is more amenable to Monte Carlo simulation. 
We construct a \emph{semidirect Ising-XY model} with Hamiltonian
\begin{equation}
  \label{eq:lattice-action}
H = \sum_{i,\mu} \left[ K_\mu \cos\left( \theta_{i+\hat{\mu}} - \theta_i - k^*_\mu \sigma_i + \delta_\mu \right) + J \sigma_{i+\hat \mu} \sigma_i \right].
\end{equation}
Each site \(i\) of the square lattice hosts degrees of freedom \(\theta_i \in [0,2\pi)\) and $\sigma_i \in \{-1,1\}$, and $\mu \in \{ x,y\}$ represents the two dimensions of the system.
$K_\mu$ is the generally anisotropic XY coupling strength, $J$ is the Ising coupling strength, and $k^*_\mu$ are the components of the wavevector $k^*$, which we take to be $|k^*| \hat x$.
We set the inverse temperature $\beta$ to unity by absorbing it into $K_\mu$ and $J$.
Finally, $\delta_\mu$ represents a global twist of the system which is equivalent to an arbitrary twist in the boundary condition by a gauge transformation.
(Note that this model differs from the Ising-XY models described in Refs.~\onlinecite{li1994,granato1991,hasenbusch2005,lee1985} which have symmetry groups different from $\rU(1) \rtimes \ZZ_2$.)
We sample from the Boltzmann distribution corresponding to Eq.~\ref{eq:lattice-action} using a Markov chain Monte Carlo algorithm.
We use a hybrid method consisting of a series of Metropolis sweeps~\cite{metropolis1953} and a cluster algorithm.
The Metropolis step consists of a sweep of the XY field followed by a sweep of the Ising field and the cluster algorithm is a modified Wolff cluster algorithm~\cite{wolff1989} with a ghost spin prescription~\cite{dotsenko1991, *rossler1999} (see End Matter for details).

To numerically identify the BKT vortex confinement transition, we measure the helicity modulus, which is the second-order response in $\delta_\mu$,
\begin{equation}
  \label{eq:helicity-modulus-def}
  \Upsilon_\mu \equiv \frac{1}{K_\mu} \left. \frac{\partial^2 f}{\partial \delta_\mu^2} \right|_{\delta_\mu=\delta_\mu^*}.
\end{equation}
Here, $\delta_\mu^*$ is the twist that minimizes the free energy density $f$. Note that the generically non-zero value of $\delta_\mu^*$, which would typically vanish in conventional superfluids, is a consequence of the interplay between the $\ZZ_2$ symmetry breaking and the coupling to the $U(1)$ phase in a finite system.
In the uncoupled model ($k^*=0$), the minimizing twist $\delta^*$ is zero by $\theta \mapsto -\theta$ symmetry, provided that $K_\mu >0$.

In the long-wavelength limit, the XY degrees of freedom in Eq.~\ref{eq:lattice-action} are described by a two dimensional anisotropic $\mathrm{O}(2)$ nonlinear sigma model, with generically different helicity moduli (couplings) in the $x$ and $y$ directions.
By rescaling the coordinates of this field theory, we obtain an equivalent isotropic field theory with a helicity modulus
\begin{equation}
  \label{eq:geometric-mean}
  \Upsilon = \sqrt{\Upsilon_x \Upsilon_y}.
\end{equation}
In the thermodynamic limit, the BKT renormalization analysis of this isotropic theory predicts that the helicity modulus jumps from $2/(\pi K)$ at the transition to zero, leading to a deconfined phase.

\begin{figure}[t]
\centering
\includegraphics[width=\columnwidth]{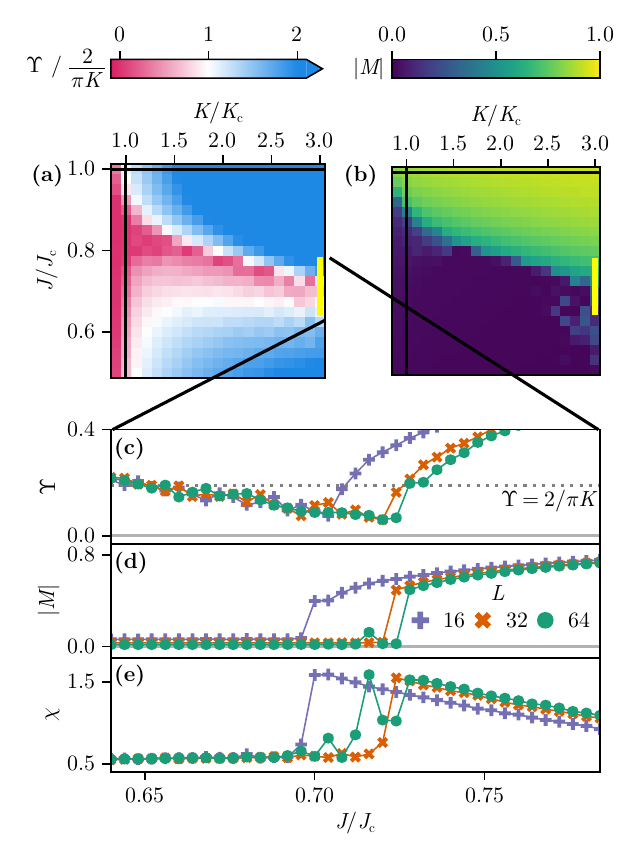}
\caption{\label{fig:phase-diagram} Monte Carlo phase diagram demonstrating vortex deconfinement near the Ising transition and a fluctuation-driven first order phase transition.
  \textbf{(a)} Helicity modulus and \textbf{(b)} magnetization for XY and Ising couplings ($K$ and $J$ respectively), scaled by their uncoupled critical values $K_\text{c}$ and $J_\text{c}$, marked with black lines.
  The BKT transition occurs at $\Upsilon = 2 / \pi K$ which is shown in white.
  \textbf{(c)} Helicity modulus $\Upsilon$, \textbf{(d)} magnetization $M$ and \textbf{(e)} magnetic susceptibility $\chi$ over a vertical cut at $K/K_\text{c} = 3$, shown as yellow lines in (a).
  The dotted line shows the universal BKT jump $\Upsilon = 2/\pi K$.
\emph{Parameters}: $k^*=0.1 \pi$, \(\delta_y=0\), \(\delta_x\) determined by minimizing the free energy using the modified Newton's method Eq.~\ref{eq:newtons-method}.}
\end{figure}%

The numerical results for the helicity modulus and Ising magnetization are shown in Fig.~\ref{fig:phase-diagram}.
Calculation details can be found in the End Matter and additional results can be found in the Supplementary Material.
Near the Ising transition, the helicity modulus drops below the universal jump of $\Upsilon = 2/\pi K$, indicating vortex deconfinement.
This stiffness collapse appears to occur even at arbitrarily large $K$ values, with a narrow region of deconfinement always accompanying the Ising transition.

Additionally, at large $K$ the Ising transition becomes first order and the susceptibility ceases to diverge.
Since a first order transition features coexistence, it is difficult to converge on the optimal twist $\delta^*_x$ near the transition, leading to noise in the magnetic susceptibility in Fig.~\ref{fig:phase-diagram}(e). 
Note that our symmetry defect argument implies that there cannot be a direct continuous transition between the ferromagnetic and paramagnetic Ising phases while vortices remain confined. 
It does not exclude a first order transition, though the mechanism for the transition should not involve percolating domain walls.\footnote{A direct first order transition between the two vortex confined phases is also not excluded, but seems implausible as such a transition is expected to be in the Ising universality class.}
Nonetheless, in the following section, we argue that the first order transition is driven by vortex deconfinement.

\section{Continuous field model}%
To further explain the stiffness collapse and first order transition in Fig.~\ref{fig:phase-diagram}, we consider a continuous field theory and use a variational method to estimate the free energy.
This theory models the Ising order parameter using a $\phi^4$ model and the XY phase as a compact field $\theta$ in which vortices are topological defects:
\begin{multline}
  \label{eq:continuum-lagrangian}
H = \int d^2 r \; \left[ \frac K 2 (\nabla \theta - k^* \phi\,\hat{x} + \vec \delta)^2 \right. \\ \left. + \frac a 2 \phi^2 + \frac 1 2 (\nabla \phi)^2 + \frac g 4 \phi^4 \right].
\end{multline}
The $\theta$ field contains both a Gaussian spin wave part and a vortex part.

The key variational principle is the Jensen-Feynman inequality~\cite{jensen1906, feynman2018, bogoliubov1958} which we apply independently to each vortex configuration $\{V\}$.
The inequality gives a bound on the partition function $Z$ which states that
\begin{equation}
  \label{eq:jensen}
Z \geq Z_\text{var} \equiv \sum_{\{V\}}  e^{-f_0 L^2 - \langle H - H_0\rangle_0 }.
\end{equation}
Here $Z_\text{var}$ is the variational partition function parameterized by an ansatz Hamiltonian $H_0$ which in turn defines the free energy density $f_0$ and the expectation value $\langle \cdots \rangle_0$.
Note that both $f_0$ and the expectation value treat the vortex configuration as fixed. 

We choose a Gaussian ansatz for the variational Hamiltonian $H_0$, which is quadratic in both fields $\phi$ and $\theta$ (excluding vortices $V$) with translationally invariant variational couplings.
This choice yields an ansatz $H_0$, which is identical to $H$ (Eq.~\ref{eq:continuum-lagrangian}) with the variable $a$ replaced by an effective mass $M$, the field $\phi$ shifted by a magnetization $\bar\phi$, and the quartic term removed. 
An explicit form is found in the End Matter.
For simplicity, we assume that the spin wave parameters are equivalent since Eq.~\ref{eq:continuum-lagrangian} does not contain higher-order terms in $\theta$.
The two variational parameters $\bar\phi$ and $M$ correspond to the order parameter and the fluctuation strength respectively.
Though this ansatz is quadratic in $\phi$, it retains the vortex contribution which is generally unsolvable.
Later we will approximate the vortex free energy using the standard nonanalytic form given by the renormalization theory of the BKT transition~\cite{kosterlitz1974,*kosterlitz2016}.

To evaluate $f_0$ and $\langle H- H_0 \rangle_0$, we integrate out the spin wave part of the $\theta$ field, leaving only the vortex degrees of freedom. We then use Wick's theorem to evaluate the $\phi^4$ term.
By performing a shift in the Ising field proportional to the vortex potential, the Ising and vortex parts in $H_0$ decouple,
leading to an effective free energy density $f_\text{var}$ given by
\begin{align}
  f_\text{var} &=  -L^{-2} \ln Z_\text{var} \nonumber \\
  &=f^\phi_\text{var} + f^V_\text{var}. \label{eq:fvar}
\end{align}
This expression has an Ising contribution independent of the vortex configuration
\begin{equation}
  \label{eq:fphivar}
  f^\phi_\text{var} = f_M + \frac{a -M + 3g\bar\phi^2}{2} \dphi + \frac {3g} 4 \dphi^2 + \frac a 2 \bar\phi^2 + \frac g {4} \bar\phi^4
\end{equation}
 where $f_M = -(1/2L^2) \sum_k \ln    G_M(k)$ is the Gaussian contribution and $S=\sum_k G_M(k)=G_M(x=0)$ is a two-point correlator which depends on a UV cutoff $\Lambda$\footnote{One can absorb the cutoff $\Lambda$ into dimensionless parameters $a\rightarrow a/\Lambda^2$, $M\rightarrow M/\Lambda^2$, $g\rightarrow g/\Lambda^2$ and $k^* \rightarrow k^*/\Lambda$.}. (See the End Matter for an explicit expression.) We omit a constant term from the spin-wave integration which depends only on $K$ and $\Lambda$.

 The vortex contribution $f^V_\text{var}$ to the free energy density is, at lowest order in the vortex fugacity, equal to that of a neutral Coulomb plasma with an anisotropic coupling constant. 
 We approximate this free energy using the known form of the Coulomb plasma free energy from the renormalization group (RG) analysis of Refs.~\onlinecite{kosterlitz1974,kosterlitz2016}.
To apply the formula, we first remove the anisotropy by rescaling the $x$ and $y$ directions (as with the helicity modulus) to obtain an effective isotropic coupling 

\begin{equation}
 \label{eq:Keff}
 \Keff = K \sqrt{1-\frac{\Kk}{M+\Kk}} + \frac{\Delta}{2} \frac{K^2 {k^*}^2}{\sqrt{M (M+\Kk)^3}}
\end{equation}
where $\Delta = a -M + 3 g (\bar\phi^2+\dphi)$. In the zero fugacity limit, the free energy is minimized by $\Delta=0$ and $\Keff$ becomes
\begin{equation}
  \Keff = K\sqrt{1-\Kk \chi}
\end{equation}
where $\chi = (M+\Kk)^{-1}$ is the magnetic susceptibility of the ansatz. This demonstrates that the effective vortex coupling is suppressed by the presence of domain walls. It also matches the results of a direct calculation of the helicity modulus from Eq.~\ref{eq:continuum-lagrangian}.

The effective coupling $\Keff$ enters the variational free energy density via the Coulomb gas free energy which near the transition has the standard approximate form~\cite{kosterlitz2016} 
\begin{equation}
  \label{eq:FCG}
  f^V_\text{var} \approx - C \exp\left( - \frac{2\pi}{\sqrt{\Keff^{-1}-K^{-1}_\text{c}}} \right) \Theta(K_\text{c} - \Keff).
\end{equation}
where $K_\text{c}$ is the BKT critical coupling---taken to be the lattice XY value of $K_\text{c} \approx 1.12$~\cite{komura2012}, $C>0$ is a proportionality constant, and $\Theta$ is the unit step function.
In the case that the coupling is negative in any direction, the vortices proliferate and the free energy is determined by the UV cutoff; this effectively sets $\Keff=0$. 
Though this is a rough scaling form given by the RG flow equations, the variational calculation is mostly sensitive to the high-temperature limit. 
In a true continuous system, this limit has a free energy given by the Debye-H\"{u}ckel approximation which diverges logarithmically in $T$ \cite{solla1981}.
However on a lattice, there is a maximum to the free energy since each lattice plaquette can only host one vortex. This motivates the form of Eq.~\ref{eq:FCG} which has a minimum free energy density of $-C$.
In fact, the presence of a first-order transition depends not on the precise form of $f^V_\text{var}$ but generally on a sufficiently sharp drop in the free energy to a constant value as $\Keff$ approaches zero.

\begin{figure}[t]
\centering
\includegraphics[width=\columnwidth]{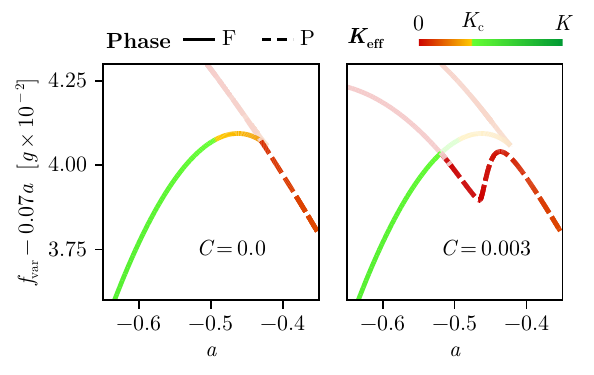}
\caption{\label{fig:free-energy}
  Variational free energy density $f_\text{var}$ in units of $g$ from the variational method shown for ferromagnetic (F) and paramagnetic (P) states for two values of the vortex free energy prefactor $C$.
  The effective vortex coupling $\Keff$ (Eq.~\ref{eq:Keff}) is shown compared to the critical value $K_\text{c}$ differentiating confinement and deconfinement.
  The unrealized states are displayed in faded colors.
  Though the $C=0$ exhibits a small discontinuity in the derivative, this point does not satisfy the Ginzburg criterion (see Fig.~\ref{fig:ginzburg-criterion}) and the variational approximation is unreliable.
  $K=3$, $k^*=1$, $g=2$, $\Lambda=1$. }
\end{figure}

We calculate the minima of $f_\text{var}$ by taking the derivative in terms of the two variational parameters--- $\bar\phi$ and $M$--- and by considering the nonanalytic point at $\Keff=0$.
Minimizing over $\bar\phi$ gives three solutions---one paramagnetic and two ferromagnetic, with opposite magnetizations.
The free energy densities of these states are plotted in Fig.~\ref{fig:free-energy} with and without the vortex contribution by setting $C$ to zero. To more easily calculate the free energy densities, we vary the mass $M$ and calculate the corresponding $a$ at which $M$ is the minimizer. Details can be found in the End Matter and Supplementary Material.

Though both cases show a discrete jump in the magnetization, only the $C>0$ case has a discontinuity in the order parameter within the range of validity of the Gaussian ansatz. 
On the ferromagnetic side, this validity can be determined by the Ginzburg criterion~\cite{ginzburg1961,levanyuk1959,amit1974}:
\begin{equation}
  \label{eq:ginzburg-criterion}
  \bar\phi^2 \gg \dphi.
\end{equation}
This inequality expresses that the order parameter's mean is far larger than its fluctuations in a symmetry breaking phase. Its violation indicates that there could be a nearby continuous phase transition, and that the Gaussian ansatz is not valid.
In the $C=0$ case, the values of \(\bar\phi^2\) and $\dphi$ are of similar order while for the $C=0.02$ case the free energy crossing occurs at a point where the order parameter is larger than the fluctuations (see Fig.~\ref{fig:ginzburg-criterion} in the End Matter).
It is worth noting for comparison that a variational treatment of the continuous transition in the 2D $\phi^4$ model always exhibits a small discontinuity in the order parameter, but that the Ginzburg criterion fails to be met near the discontinuity.

The variational results demonstrate that the vortex contribution can lead to a fluctuation-driven first order transition from a ferromagnetic-confined state to a paramagnetic-deconfined state.
The fluctuations in the Ising field drive the transition since domain walls reduce the effective vortex interaction and in turn lead to a paramagnetic state.
Such an effect is analogous to the fluctuation-driven first order transition in Ref.~\onlinecite{halperin1974} in which a mediating magnetic field leads to a cubic term in the Ginzburg-Landau free energy, generating a free energy crossing.
However, in the semidirect Ising-XY model the mediating vortex field depends on the susceptibility of the Ising field, not on the order parameter, which precludes a pure mean-field treatment as in Ref.~\onlinecite{halperin1974}.

\section{Discussion}%
Both numerical simulation and theoretical analysis predict that the phase diagram of the semidirect Ising-XY model has two disconnected confined phases for any \(k^* \not\in 2\pi \mathbb{Z}\).
Nonetheless, the numerical results cannot definitively verify the existence of a separating deconfined region in the vicinity of the Ising transition at large vortex stiffness, as predicted by our theoretical symmetry defect argument.
Additionally, the variational argument supports the numerical evidence 
that the BKT transition merges with the Ising transition and becomes a fluctuation-driven first order transition.
The first order transition implies a tricritical point which occurs around $J \approx 0.8 J_\text{c}$ and $K \approx 1.5 K_\text{c}$ in Fig.~\ref{fig:phase-diagram}.
It is still an open question as to whether the BKT transition merges with the Ising transition exactly at this point or if the triple point and the tricritical point are distinct.

The lattice model (Eq.~\ref{eq:lattice-action}) reflects the phenomenology of the Bose gas with a double-well dispersion, where one associates the parameters of the former--- $J,K,k^*$ and the lattice constant $a$ (which is implicitly set to $a=1$ by choice of units)--- with parameters of the latter.
We outline this correspondence for the spin-orbit coupled Bose gas (analogous correspondences can be drawn in the various other realizations of double-well dispersions~\cite{galitski2013,zheng2014,atala2014observation}).
In this case, $k^*$ is the momentum difference between the two spin components of the Bose gas.
The parameter $K$ is proportional (up to factors of $a$) to the superfluid stiffness $\hbar^2 n_\text{2D}/m_\text{B}$ of the two dimensional Bose gas with density $n_\text{2D}$ and mass $m_\text{B}$.  
The parameter $J$ is determined by the Ising domain-wall energy~\cite{liu2016exotic} and $a$ can be associated with the Ising domain-wall width of the two dimensional spin-orbit coupled Bose gas. 
This dictionary allows the predicted phase diagram of Fig.~\ref{fig:cartoon} to be translated to the bosonic models of our original interest.

This work opens the door to future paths of research.
For example, the phase diagram for the 3D semidirect Ising-XY model is unknown. 
In this case, the XY spins feature true long-range order with line defects.
Additionally, it is unknown if there is a deeper relationship between phase diagram topology and non-Abelian symmetry groups.
For example, lattice models can feature Euclidean lattice symmetry groups which are the semidirect product of translations and rotations.
Previous work \cite{manjunath2021} on crystalline gauge fields demonstrated that defects in this lattice symmetry can bind charge, which is reminiscent of vortex charges bound to Ising domain walls. 
Do these observations constrain the possible transitions between spatial symmetry breaking phases? We leave this question for future work.

\begin{acknowledgments}
   We thank Sayak Bhattacharjee and G. J. Sreejith for valuable discussion.
   This work is supported by the Laboratory for Physical Sciences, through their support of the Condensed Matter Theory Center at the University of Maryland.
   SYT thanks the Joint Quantum Institute at the University of Maryland for support through a JQI fellowship. 
   DML is additionally supported by a Stanford Q-FARM Bloch Fellowship, and a Packard Fellowship in Science and Engineering (PI: Vedika Khemani).
\end{acknowledgments}%

\bibliography{bib.bib}

\clearpage
\appendix

\setcounter{equation}{0}
\setcounter{figure}{0}
\makeatletter
\renewcommand{\theequation}{A\arabic{equation}}
\renewcommand{\thefigure}{A\arabic{figure}}

\oldsection{End Matter}
\section{Monte Carlo Details}%
Here we describe details of the Monte Carlo simulation. We outline the cluster algorithm, give explicit expressions for the helicity modulus, and describe the modified Newton's method.

The Wolff cluster algorithm~\cite{wolff1989} reduces the critical slowing down near phase transitions by flipping large domains of spins at once.
We can use it to reduce the presence of metastable states characterized by Ising domains.
It consists of recursively growing a spin cluster from a single site and then flipping the entire cluster.
To apply the algorithm to our system we begin by mapping the entire model onto an Ising model.
We assume $\delta_y=0$.
We define $\eta_i = \sigma_i (\theta_i - e)$ where $e$ is a random constant which reflects the $\mathrm{SO}(2)$ symmetry of $\theta$.
Applying this decouples the Ising degree of freedom from the XY spin:
\begin{equation}
H = \sum_{\langle ij\rangle}  J_{ij} \sigma_i \sigma_j + \sum_i h_i \sigma_i + C[\eta]
\end{equation}
where
\begin{align*}
  J_{ij} &= J +\begin{cases}
    K_x \cos \delta_x \; \sin \eta_j \; \sin\left( \eta_i + k^* \right) & \vec r_{ji} = \hat x \\
    K_x \cos \delta_x \; \sin \left( \eta_j + k^* \right)  \; \sin \eta_i  & \vec r_{ji} = -\hat x \\
    K_y \sin \eta_j \; \sin \eta_i & \vec r_{ji} = \pm \hat y,
  \end{cases} \\
h_{i} &= K_x \sin \delta_x  \left[ \sin \eta_i  \cos (\eta_{i-\hat x} + k^*)  \right. \nonumber \\ & \hspace{4cm} \left. - \sin (\eta_i+k^*) \cos \eta_{i+\hat x} \right]
\end{align*}
and $C[\eta]$ is a constant in $\sigma$.
The presence of a global twist $\delta_x$ results in an effective magnetic field.
This can be incorporated into the cluster algorithm as a ``ghost spin''~\cite{dotsenko1991, rossler1999} which acts as a static spin, converting the background field to an Ising interaction from this ghost spin to every other spin.
The cluster algorithm can then be performed on this augmented graph. If the ghost spin is included in the cluster, the algorithm is aborted.

The helicity modulus for the Hamiltonian in Eq.~\ref{eq:lattice-action} as defined by Eq.~\ref{eq:helicity-modulus-def} is
\begin{equation}
  \label{eq:helicity-modulus}
  \Upsilon_\mu = \langle e_\mu \rangle - K_\mu L^2 \; \mathrm{Var}\left[ s_\mu \right]
\end{equation}
where
\begin{align*}
  e_\mu &= L^{-2} \sum_i \cos(\theta_{i+\hat{\mu}} - \theta_i - k^*_\mu \sigma_i + \delta_\mu ) \\
  s_\mu &= L^{-2} \sum_i \sin(\theta_{i+\hat{\mu}} - \theta_i - k^*_\mu \sigma_i + \delta_\mu ).
\end{align*}
for a $L\times L$ system.

The Monte Carlo simulation consists of a series of optimization steps to find the optimal twist $\delta_x^*$ minimizing the free energy followed by a measurement simulation.
For each optimization step, we perform a full Monte Carlo simulation consisting of $10^7$ thermalizing sweeps and $10^7$ measurement sweeps, applying the cluster algorithm every 10 sweeps and measuring every 100.
Since $\delta_x$ lies on a compact manifold $[0,2\pi/L_x)$, we use a modified Newton's method
\begin{equation}
  \label{eq:newtons-method}
          \delta_x \mapsto \delta_x - \frac{1}{L_x} \tan^{-1}\left( \frac{\partial f}{\partial \delta_x},\;  \frac{\partial^2 f}{\partial \delta_x^2}  \right)
\end{equation}
where $\tan^{-1}(y,x)$ is the two-argument inverse tangent and $\partial f / \partial \delta_x = - K_x \left\langle s \right\rangle$.

The large number of measurements is necessary to accurately compute the second derivative in Eq.~\ref{eq:newtons-method}, since the error scales with the square root of the samples size due to the dependence on the variance (see Eq.~\ref{eq:helicity-modulus}).
Furthermore, the variance is particularly sensitive to autocorrelations, justifying the large measurement period.
Nonetheless, using the second derivative over typical gradient descent helps avoid slow convergence when starting near a maximum of the free energy.
We initialize in a cold start (that is, in a ground state), reinitializing for each gradient descent step, and iterate until the difference in the measured free energy density $f$ between subsequent gradient descent steps is less than $10^{-7}$. At this point an additional Monte Carlo simulation gives the observables plotted in Fig.~\ref{fig:phase-diagram}.

\section{Detailed variational calculation}
We begin deriving Eqs.~\ref{eq:fvar} and \ref{eq:Keff} by integrating out the spin wave part of $\theta$ in both the 
Hamiltonian of Eq.~\ref{eq:continuum-lagrangian} and the ansatz Hamiltonian.
To perform this transformation, we decompose the gradient of the field into a spin wave part and vortex part: $\nabla \theta = \nabla \varphi - \nabla \times (\hat z \psi )$ where $\varphi$ is a scalar field and $\psi(r)=\sum_i V_i \ln|r-r_i|$ is potential from the vortex field.
Separating the degrees of freedom converts the Hamiltonian to $H = H^\phi + H^V$
where 
\begin{align}
H^\phi &= \int d^2 r \left( \frac K 2 [ \nabla \varphi - k^* (\phi - \langle \phi \rangle)\,\hat{x}]^2 \right. \nonumber \\ 
&\hspace{3cm} \left. + \frac a 2 \phi^2 + \frac 1 2 (\nabla \phi)^2 + \frac g 4 \phi^4 \right) \\
H^V &= \frac{K}{2} \int d^2 r\; \left[ (\nabla \psi)^2 + 2 k^*  \psi \partial_y \phi\right].
\end{align}
We have set the twist $\vec \delta$ to the value $k^* \langle \phi \rangle \hat x$ which minimizes the free energy.

Integrating out the spin-wave field $\varphi$ gives
\begin{multline}
  \label{eq:Hphi}
\frac{H^\phi}{L^2} = \frac{1}{2} \sum_{k} \;   G_a^{-1}(k) |\phi_k|^2 + \Kk \left( \frac{\langle\phi\rangle}{2} - \phi_0 \right) \langle\phi\rangle \\ + \frac g {4L^2} \int dx\, \phi^4
\end{multline}
where $  G_m(k)$ is the renormalized propagator, given by
\begin{equation}
    G^{-1}_m(k) = k^2 + m + K{k^*}^2 \left( 1 - \delta_{k0} \frac{k_x^2} {k^2} \right)
\end{equation}
for arbitrary mass $m$.
We omit an extra term which depends only on $K$.
The variational ansatz is then $H_0=H_0^\phi + H^V$ where
\begin{equation}
H_0^\phi  = \frac{L^2}{2} \sum_{k}   G^{-1}_M(k) |\delta \phi_{k}|^2 \label{eq:variational-1}
\end{equation}
and $\delta\phi = \phi - \bar\phi$.

We can remove the coupling between $\psi$ and $\delta\phi$ in $H$ by shifting $\delta\phi$ to%
\begin{equation}
  \xi \equiv \delta \phi + \chi \hspace{0.5cm} \text{where} \hspace{0.5cm} \chi_k = iK{k^*}  k_y   G_M(k) \psi_k.
\end{equation}
Substituting the shifted field decouples the Hamiltonian, giving
\begin{equation*}
\frac{H_0}{ L^{2}} = \frac{1}{2} \sum_{k} \left[   G^{-1}_M(k) \left( |\xi_k |^2 - |\chi_k|^2 \right)  + K k^2 |\psi_k|^2 \right]
\end{equation*}
which has a free energy density
\begin{align}
  \label{eq:f0}
 f_0 %
 &= f_M + \frac{1}{2} \sum_{k}  \left[ K  k^2 |\psi_k|^2 -   G_M^{-1}(k) |\chi_k|^2  \right].
\end{align}
Here, $f_M = -(1/2L^2) \sum_k \ln   G_M(k)$ is the Gaussian free energy density which can be approximated by an analytically-solvable integral over momenta giving
\begin{equation}
f_M  \approx \frac{1}{8\pi^2} \int_\Lambda d^2 k \, \ln   G^{-1}_M(k) = \frac{I(\Lambda^2+M) - I(M)}{8\pi} 
\end{equation}
where %
\begin{multline*}
  I(u) = (\Kk + 2u) \ln\left( \sqrt{\Kk + u} + \sqrt{u} \right) \\ - \sqrt{u(\Kk+u)} - u \ln 4.
\end{multline*}

Next we evaluate $\langle H-H_0\rangle_0 = \langle H^\phi - H^\phi_0 \rangle_0$. First, we shift the Ising field $\phi$ to the field $\xi$ in the Hamiltonian of Eq.~\ref{eq:Hphi}:
\begin{multline*}
  \frac{H^\phi}{ L^{2}} = \frac 1 2 \sum_k \left[  G^{-1}_a(k) + 3g\bar\phi^2\right] \left( |\xi_k|^2 + |\chi_k|^2 \right)  \\
  + \frac a 2 \bar\phi^2 + \frac{g}{4} \bar\phi^4 
  + \frac{g}{4L^2} \int dx\; \left( \xi^4 + 6\xi^2 \chi^2  \right) + \cdots
\end{multline*}
where we omit terms that are odd in $\xi$ or $\phi$ as well as the $ \chi^4 $ term, which for sufficiently small vortex fugacity is negligible.

Taking the expectation value and using Wick's theorem, the difference becomes
\begin{multline}
  \label{eq:difference}
  \frac{\langle H - H_0 \rangle_0 }{L^2} =
   \frac{a -M + 3g\bar\phi^2}{2} \dphi + \frac {3g} 4 \dphi^2 \\ + \frac \Delta 2 \sum_k  |\chi_k|^2 + \frac a 2 \bar\phi^2  + \frac{g}{4} \bar\phi^4 
\end{multline}
where 
$\Delta = a -M + 3 g (\bar\phi^2+\dphi)$
and
the fluctuation strength $\dphi = \langle \xi^2\rangle_0$ evaluates to
\begin{align*}
  \dphi &\approx \int_\Lambda \frac{d^2 k }{(2\pi)^2}    G_M(k) \nonumber \\
   &= \frac{1}{2\pi} \ln \left( \frac{\sqrt{K{k^*}^2+M+\Lambda^2} + \sqrt{M+\Lambda^2}} {\sqrt{K{k^*}^2+M}+ \sqrt{M}} \right).
\end{align*}

Summing Eqs.~\ref{eq:f0} and \ref{eq:difference} results in a term $f^\phi_\text{var}$ (see Eq.~\ref{eq:fphivar}), which is independent of the vortex field, and a vortex Hamiltonian $H^V_\text{eff}$, defined as
\begin{equation}
 H^V_\text{eff} = \frac 1 2 \sum_k \mathfrak K(k) |\psi_k|^2
\end{equation}
where
\begin{equation}
 \mathfrak K (k) = K\left(k^2 - \Kk k_y^2 G_M(k)\left[1 - \Delta G_M(k)\right] \right).
\end{equation}
Substituting the vortex density $V_k = (k^2/2\pi) \psi_k$ and Ising propagators, the effective Hamiltonian becomes
\begin{equation}
 H^V_\text{eff} = 2\pi^2 K \sum_k \frac{M^2 k^2 + \Kk (M + \Delta) k_y^2}{\left(M k^2 + \Kk k_y^2\right)^2} |V_k|^2.
\end{equation}
By rescaling $k_x^2 \to k_x^2 (\Kk + M) / M$, we obtain two terms: a $k^{-2}$ term and $(k_y^2 - k_x^2)/k^4$ term, the latter of which results in a $\sim x^2 / r^2$ real space coupling. Since this second term is irrelevant to the RG and the BKT transition, we ignore it and obtain the effective XY coupling found in Eq.~\ref{eq:Keff}

\section{Ginzburg criterion}%
\begin{figure}[t]
\centering
\includegraphics[width=1.0\columnwidth]{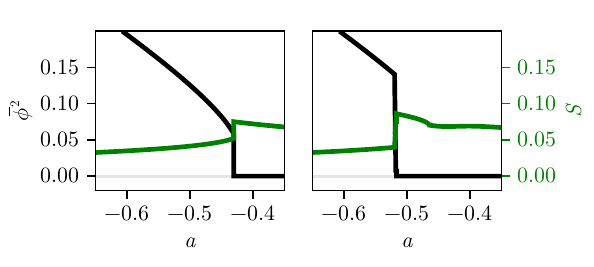}
\caption{\label{fig:ginzburg-criterion}
  The order parameter $\bar \phi^2$ compared with the fluctuation strength $\dphi$ for the variational state that minimizes the free energy in Fig.~\ref{fig:free-energy}, demonstrating the approximate satisfaction of the Ginzburg criterion (Eq.~\ref{eq:ginzburg-criterion}) when $C>0$ since the $\bar\phi^2$ is around 3 times larger than $\dphi$.
}
\end{figure}
The order parameter compared with the fluctuation strength gives a measure of the validity of the variational approximation in Fig.~\ref{fig:free-energy} according to the Ginzburg criterion (Eq.~\ref{eq:ginzburg-criterion}).
We compare these two quantities in Fig.~\ref{fig:ginzburg-criterion} to argue that the discontinuity in the derivative of the free energy indicates a true first order transition.

\let\section\oldsection
\newpage
\appendix 
\onecolumngrid

\clearpage
\renewcommand\thefigure{S\arabic{figure}}    
\setcounter{figure}{0} 
\setcounter{page}{1}
\renewcommand{\theequation}{S\arabic{equation}}
\setcounter{equation}{0}
\renewcommand{\thesubsection}{SM\arabic{subsection}}
\thispagestyle{empty}

\section{\LARGE Supplementary Information}

\newcommand{\bKeff}{K_\text{eff}}
\newcommand\FV{f_\text{var}}

\section{Additional Monte Carlo results}%
\FloatBarrier
Here we present additional data from the Monte Carlo simulation, including a scaling analysis of the first order magnetic transition, the directional helicity moduli $\Upsilon_x$ and $\Upsilon_y$, and the magnetic susceptibility.

\begin{figure}
\centering
\includegraphics[width=0.6\columnwidth]{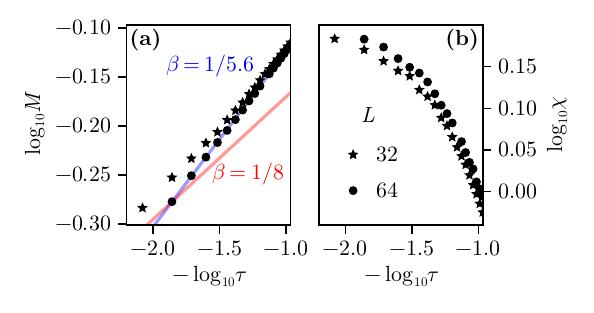}
\caption{\label{fig:scaling-exponents}
  Near-critical scaling of \textbf{(a)} magnetization $M$ and \textbf{(b)} susceptibility $\chi$ as a function of reduced temperature $\tau \equiv J^* / (J - J^*)$ where $J^* \approx 0.722$ at $K = 3K_\text{c}$.
  The magnetization is fit to the exponent $\beta=1/5.6$ before the first order transition, which differs from the 2D Ising scaling exponent $\beta=1/8$.
}
\end{figure}

To analyze the scaling around the magnetic transition, we plot the magnetization and susceptibility on a logarithmic scale around the transition in Fig.~\ref{fig:phase-diagram}(c-e). 
These plots are shown in Fig.~\ref{fig:scaling-exponents}.
Though the magnetization appears to scale as a power law before the first order transition, the susceptibility does not apparently diverge, saturating at a fixed value.

\begin{figure}
\centering
\includegraphics[width=0.8\columnwidth]{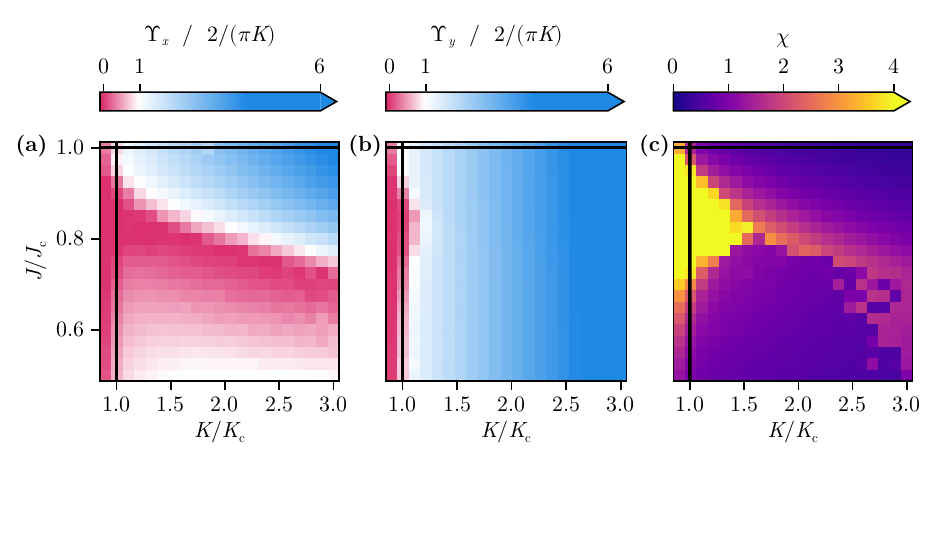}
\caption{\label{fig:extra-monte-carlo}
  Additional Monte Carlo results.
  \textbf{(a)} $x$-direction helicity modulus,
  \textbf{(b)} $y$-direction helicity modulus, and
  \textbf{(c)} magnetic susceptibility.
The parameters are the same as those from Fig.~\ref{fig:phase-diagram}(a-b) in the main text.
}
\end{figure}

In Fig.~\ref{fig:extra-monte-carlo}, we plot the helicity moduli $\Upsilon_x$ and $\Upsilon_y$ as well as the magnetic susceptibility $\chi$. 
The results show that the vortex deconfinement around the Ising transition is predominantly driven by the horizontal stiffness $\Upsilon_x$ while the vertical stiffness $\Upsilon_y$ is relatively independent of $J$. 
Additionally, we see the divergence in the susceptibility disappears for $K/K_\text{c} \geq 1.6$.


\section{Finding the minimum of \texorpdfstring{$f_\text{var}$}{fvar}}
We can minimize the variational free energy by deriving it with respect to our two variational parameters:
We begin by taking the derivatives of the effective vortex coupling $\bKeff$:
\begin{align}
  \frac{1}{K} \frac{\partial \Keff}{\partial \bar\phi^2}  &= \frac{3}{2} \frac{g \Kk}{\sqrt{M(M+\Kk)^3}} \\
  \frac{1}{K} \frac{\partial \Keff}{\partial M}  &=  - \Kk \; \frac{3g M (M+\Kk) \Pi_0 + \frac{1}{2}\Delta(\Kk + 4M)}{2\sqrt{M^3(M+\Kk)^5}}
\end{align}
where
$\Pi_0$ is the polarization operator, given by
  \begin{align}
    \Pi_0 &= \sum_k   G^2_M(k) \\
    &= \frac 1 {
4\pi} \left( \frac 1 {\sqrt{M (M+ K{k^*}^2)}} - \frac 1 {\sqrt{(\Lambda^2 + M)(\Lambda^2 + M + K{k^*}^2) }} \right).
  \end{align}
The free energy density derivatives are then
\begin{align}
  \frac{\partial \FV}{\partial \bar\phi}  &= \left[ a + 3g\dphi + g\bar\phi^2 + 2 \frac{\partial f^V_\text{var}}{\partial \Keff} \frac{\partial \Keff}{\partial \bar\phi^2} \right] \bar\phi \\
  &= \left[ \tilde a + 3g\dphi - 2g\bar\phi^2 + 2 \frac{\partial f^V_\text{var}}{\partial \Keff} \frac{\partial \Keff}{\partial \bar\phi^2} \right] \bar\phi \label{eq:minimum-phi} \\
  \frac{\partial \FV }{\partial M}  &= -\frac{\Pi_0}{2} \left( \tilde a-M+3g\dphi\right) + \frac{\partial f^V_\text{var}}{\partial \Keff} \frac{\partial \Keff}{\partial M} \label{eq:minimum-M}
\end{align}
where $\tilde a = a + 3g\bar\phi^2$.

We first notice that the solutions of Eq.~\ref{eq:minimum-phi} are threefold: $0$ and $\pm\bar\phi\neq 0$.
Consider the two ferromagnetic cases, both of which we parameterize without loss of generality by $\bar\phi^2$:
\begin{equation}
  \bar\phi^2 = \frac {\tilde a} {2g} + \frac{3}{2}\dphi + \frac{1}{g} \frac{\partial f^V_\text{var}}{\partial \Keff} \frac{\partial \Keff}{\partial \bar\phi^2}.
\end{equation}
Therefore finding the minimum of $M$ as a function of $\tilde a$ is sufficient to find the minimizing magnetization.

However instead of calculating the value of $M$ that solves Eq.~\ref{eq:minimum-M}, it is easier to calculate the value of $\tilde a$ that corresponds to a given $M$ since we avoid inverting $\dphi$ and $\Pi_0$ which are functions of $M$ but not $\tilde a$.
We can therefore easily find the roots of Eq.~\ref{eq:minimum-M} as a function of $\tilde a$ numerically for a range of fixed $M$ values.

When the fugacity is zero, resulting in a complete absence of vortices, the minimization equations become
\begin{align}
  \left( \tilde a + 3g\dphi \right) \bar\phi  &= 0\\
  \tilde a-M+3g \dphi &= 0
\end{align}
and the free energy is minimized by 
\begin{align}
  M &= \tilde a + 3g \dphi,
\end{align}
at which point $\Delta=0$.

\end{document}